\documentclass[a4paper,12pt]{article}

\usepackage[dvips]{graphicx}
\graphicspath{{.}}

\begin{document}
\mathsurround=2pt \sloppy
%\begin{center}
\title{\bf Parametric instability of homogeneous precession of spin in the superfluid $^{3}He-B$
}
\author{E. V. Surovtsev,  I. A. Fomin \\{\it P.L.Kapitza Institute for Physical Problems
}
\\ {\it 119334 Moscow Russia}}
%\end{center}
\maketitle

\begin{abstract}
Stability of homogeneous precession of spin due to parametric
excitation of spin waves is considered as the explanation of the
"catastrophic relaxation", that is observed in the superfluid
$^{3}He-B$. It is shown, that at sufficiently low temperatures
homogeneous precession of spin becomes unstable (Suhl
instability). At zero temperature increments of growth for all
spin wave modes are found. Estimation of the temperature of
transition to the unstable state is made.
\end{abstract}

 {\bf 1}. The use of pulsed NMR is based on the investigation of homogeneous precession of
 spin in a constant magnetic field. Spin precession induces the
 free induction signal (FIS), which is registered in the induction coil. Precession of spin in
 the superfluid B-phase of$^{3}$He has its specifics. At temperatures $T\geq 0,4T_c$, where $T_c$ is the
 temperature of superfluid transition, FIS
  exists anomalously long -- many times longer than the time
 of dephasing of spin due to a residual
 inhomogeneity of d.c. field. At $T\sim 0,4T_c$ occurs transition
 into the other regime, when on the contrary FIS disappears
 quickly. This fast decay of precession was first observed in ref.
 \cite{nyeki} and since that is referred as {\it catastrophic relaxation}.
 While anomalously long FIS was  explained a long time
 ago, even a qualitative explanation for the catastrophic
 relaxation is missing. The decay of homogeneous precession was demonstrated
 by numeric simulation of equations of spin dynamics \cite{BG}.
 The simulation was made in the restricted geometry and the authors of simulation attribute the main
 role in the destruction of precession to the walls, i.e. they
 consider the mechanism of destruction to be surface.

In the present paper  the explanation of catastrophic relaxation
is suggested, which is based on a bulk effect. It is the
instability of homogeneous precession with respect to decay into
parametrically excited spin waves with opposite wave vectors (Suhl
instability \cite{Suhl}).

Fast decay of precession was also observed in the $uudd$-phase of
solid $^{3}$He and it was explained by the onset of Suhl
instability \cite{mizus}. However, the quantitative interpretation
of the results, concerning instability of precession in the cited
work is based on the modification of theory Ref. \cite{Ohmi},
developed for the continuous NMR and  which is therefore
applicable only for small tipping angles. In our analysis this
restriction is not used and it can be applied for the arbitrary
angles between spin and magnetic field.

{\bf 2}. In order to describe motion of spin  we use expression of
Hamiltonian of Leggett, which is written in terms of  Euler angles
$\alpha$, $\beta$, $\Phi=\alpha+\gamma$ (z-axis is oriented
opposite to the direction of d.c. magnetic field {$\bf H_{0}$}) as
coordinates and canonically conjugated momenta
$P=S_{z}-S_{\zeta}$, $S_{\beta}$, $S_{\zeta}$, where $S_{z}$
--- is projection of spin onto \emph{z}-axis, $S_{\zeta}$ ---its
projection onto axis $\zeta=\hat{R}\hat{\emph{z}}$ and $S_{\beta}$
--- is projection on the line of nodes (see for details \cite{Fomin1}).
We choose units of measurement so that $\chi=g=1$, where $\chi$
--- is magnetic susceptibility per unit volume of $^{3}He-B$, and
\emph{g} --- is the gyromagnetic ratio for nuclei of $^{3}He$;
after that spin has dimensionality of frequency and energy of the
frequency squared. Using variable and units mentioned above one
can write the Hamiltonian in the form:
  \begin{equation}
     \label{Hamiltonian}
     H=\frac{1}{1+\cos{\beta}}\{S_{\zeta}^2+P
     S_{\zeta}+\frac{P^2}{2(1-\cos{\beta})}\}+\frac{1}{2}S_{\beta}^2+F_{\nabla}-\omega_{L}(P+S_{\zeta})+U_{D}(\beta,\Phi),
  \end{equation}
where $\omega_L$ - is Larmor frequency corresponding to the d.c.
magnetic field, $F_{\nabla}$ - gradient energy,$U_{D}(\beta,\Phi)$
- dipole energy. $U_{D}$ for  $^{3}He-B$ depends only on two
variables $\beta$ and $\Phi$, that justifies the choice of $\Phi$
as a variable, when  the dipole energy is essential. Gradient
energy for $^{3}He-B$ can be  written as:

\begin{equation}
   \label{F_nabla}
     F_{\nabla}=\frac{1}{2}[c_{\|}^{2}\delta_{ik}\delta_{\xi
     \eta}-(c_{\|}^{2}-c_{\bot}^{2})(\delta_{i\xi}\delta_{k\eta}+\delta_{i\eta}\delta_{k\xi})]\omega_{i\xi}\omega_{k\eta},
  \end{equation}
  where
  \begin{eqnarray}
     \label{frequency}
     \omega_{1\xi}=-\alpha_{,\xi}\sin\beta\cos\gamma+\beta_{,\xi}\sin\gamma,\nonumber \\
     \omega_{2\xi}=\alpha_{,\xi}\sin\beta\sin\gamma+\beta_{,\xi}\cos\gamma,~~\\
     \omega_{3\xi}=\alpha_{,\xi}\cos\beta+\gamma_{,\xi},{~~~~~~~~~~~~~~~}\nonumber
  \end{eqnarray}
$\alpha_{,\xi}=\frac{\partial\alpha}{\partial x_{\xi}} $ etc.,$
c_{\|}^2 {}\mbox{ è } c_{\bot}^2  $ --- are squared velocities of
two types of spin waves ("longitudinal" and "transverse"). In what
follows we choose  units in  such a way as to $c_{\|}^2=1$, then
wave vectors entering the equation will also have dimensionality
of frequency. Equations of motion, that are generated by
Hamiltonian have the form:
  \begin{eqnarray}
     \label{Ham_equations}
     \frac{\partial\alpha}{\partial t}=\frac{\partial H}{\partial P},{~~~} \frac{\partial P}{\partial t}=-\frac{\partial
     H}{\partial\alpha}+\frac{\partial}{\partial x_\xi}\frac{\partial
     H}{\partial\alpha_{,\xi}},{~~}
  \end{eqnarray}
  etc.

System of equation (\ref{Ham_equations}) has spatially uniform
stationary solution describing precession of spin in the
stationary magnetic field at
$0\leq\beta<\theta_0=\arccos(-\frac{1}{4})$ :

\begin{eqnarray}
     \label{stat_solution}
     \alpha=\omega_L t+\alpha_0,{~~}\gamma=-\omega_L
     t+\Phi^0-\alpha_0,\nonumber\\
     P^{(0)}=\omega_L(\cos\beta-1),{~~}S_\beta^{(0)}=0,{~~}S_z^{(0)}=\omega_L\cos\beta,\\
     \cos\Phi^{(0)}=(\frac{1}{2}-\cos\beta^{(0)})/(1+\cos\beta^{(0)}).\nonumber
     \
  \end{eqnarray}
  It is convenient to introduce $\psi=\alpha+\omega_Lt$
  instead of $\alpha$ and at the same time to transform the Hamiltonian
  $\tilde{H}=H+\omega_LP$, so that $\frac{\partial\psi}{\partial t}=0$.

Let us now obtain equations for the small deviations from the
stationary solution:

\begin{equation}
\delta\psi(\textbf{r},t)=\psi-\psi^{(0)},
\end{equation}
etc.

In zeroth approximation  on the small deviations the gradient
energy has three groups of terms: "stationary"$~$- with
time-independent coefficients, and two "oscillating",
corresponding to Larmor and doubled Larmor frequencies. Without
the loss of generality we consider perturbations propagating in
$yz$ plane:

   \begin{eqnarray}
      \label{F_oscill1}
      F_{\nabla
      st}=\frac{1}{2}[\delta\psi_{,y}^2[1-\mu\sin^2{\beta^{(0)}}]+(1-\mu)\delta\beta_{,y}^2+\delta\gamma_{,y}^2+2\delta\psi_{,y}\delta\gamma_{,y}\cos\beta^{(0)}+\nonumber\\
  \delta\psi_{,z}^2[1-2\mu\cos^2{\beta^{(0)}}]+\delta\beta_{,z}^2+\delta\gamma_{,z}^2(1-2\mu)+2\delta\psi_{,z}\delta\gamma_{,z}\cos\beta^{(0)}(1-2\mu)] ,\\
      \label{F_oscill2}
      F_{\nabla
      osc1}= -\mu[\delta\psi_{,y} \delta\psi_{,z} \sin
      2\beta^{(0)}\sin\omega_L t+\nonumber\\
      +(\delta\psi_{,y}\delta\gamma_{,z}+\delta\psi_{,z}\delta\gamma_{,y})\sin\beta^{(0)}\sin\omega_L t+
      (\delta\psi_{,y}\delta\beta_{,z}+\delta\psi_{,z}\delta\beta_{,y})\cos\beta^{(0)}\cos\omega_L t+\nonumber\\
      +(\delta\beta_{,y}\delta\gamma_{,z}+\delta\beta_{,z}\delta\gamma_{,y})\cos\omega_L t],\\
      \label{F_oscill3}
      F_{\nabla
      osc2}= -\frac{\mu}{2}[\sin^2{\beta^{(0)}}\delta\psi_{,y}^2\cos{2\omega_L
      t}-\delta\beta_{,y}^2\cos{2\omega_L
      t}+ \nonumber   \\
       +\sin\beta^{(0)}\delta\psi_{,y}\delta\beta_{,y}\sin{2\omega_Lt}].
   \end{eqnarray}
Oscillating terms are proportional to $\mu=1-c_{\bot}^2/c_{\|}^2$.
We assume that $\mu$ is small, and consider oscillating terms in
the equations of motion  as small perturbations. Actually $\mu$ is
not very small ($\mu\approx1/4$ in a vicinity of $T_c$), however,
this approximation gives satisfactory results. More precise
criteria of application of such approximation will be formulated
in a process of solution.

In order to use the theory of perturbation we write linearized
system of equations of motion in the form:

    \begin{equation}
          \label{V_equation}
          \frac{d\mathbf{X}}{dt}=\left(\hat{M}_0+\hat{V}(t)\right)\mathbf{X},
    \end{equation}
    where
    $$
\textbf{X}=\left(
\begin{array}{c}
\delta\psi\\
\delta\beta\\
\delta\Phi\\
\delta S_z \\
\delta S_\beta\\
\delta P
\end{array}
\right).
$$

Matrix operator $\hat{M}_0$ includes all time-independent terms,
and oscillating terms are collected in  matrix operator
$\hat{V}(t)$, which is proportional to $\mu$ and therefore is
considered as a perturbation. Equation of zero order approximation
on perturbation:
\begin{equation}
              \label{M_equation}
        \frac{d\mathbf{X}}{dt}=\hat{M}_0\mathbf{X},
    \end{equation}
    gives us dispersion laws for the three branches of spin waves:
$$
    \begin{array}{lc}
        \omega_1^2=k^2,\\
        \omega_2^2=\frac{1}{2}\omega_L^2+k^2-\frac{1}{2}\omega_L\sqrt{\omega_L^2+4k^2},\\
        \omega_3^2=\frac{1}{2}\omega_L^2+k^2+\frac{1}{2}\omega_L\sqrt{\omega_L^2+4k^2}
    \end{array}
    \eqno \stepcounter{equation}(\arabic{equation})
    \newcounter{tenth1}
    \setcounter{tenth1}{\value{equation}}
    $$
    and eigenvectors, corresponding to each oscillation branch
    $\textbf{X}_i(k),~i=1,2,3$. Recall that we are speaking about
    the spin waves propagating against the background of homogeneous
    precession, so the mentioned oscillations of spin are different from the
    usual spin waves, which are created by small deviation from
    equilibrium orientation. It is assumed here that
    $\omega_L\gg\Omega$, where $\Omega$ is the frequency of
    longitudinal oscillations.

Solution of the system (\ref{M_equation}) is given by:
\begin{eqnarray}
         \label{zero_}
         \textbf{X}(t,k,y,z)=\sum^3_{i=1} \{ a_{i-}\textbf{X}_i(k)\exp(-i\omega_i(k)t+i\textbf{k}\textbf{r} )\nonumber\\
                       a_{i+}\textbf{X}_i^{*}(k)\exp(i\omega_i(k)t+i\textbf{k}\textbf{r})+\nonumber\\
                       a_{i-}^{*}\textbf{X}_i^{*}(k)\exp(i\omega_i(k)t-i\textbf{k}\textbf{r})+\nonumber\\
                       a_{i+}^*\textbf{X}_i(k)\exp(-i\omega_i(k)t-i\textbf{k}\textbf{r}) \},
    \end{eqnarray}
where $a_i$ -  are constant coefficients, $\textbf{k}=(k_y,k_z)$
and $\textbf{r}=(y,z)$ are 2D vectors. When $\hat{V}(t)$ is taken
into account $\textbf{X}(t,k,y,z)$, given by equation
(\ref{zero_}), does not satisfy the equation of motion
(\ref{V_equation}). The first order approximation on $\hat{V}$ can
be obtained by the method of averaging of the classical mechanics.
Substituting zero order approximation (\ref{zero_}) into formulas
(\ref{F_oscill2}),(\ref{F_oscill3}) we can see that terms of the
first order  on $\mu$ will not vanish after time averaging only if
there are resonance relations between $\omega_L$ and
eigenfrequencies $\omega_i(k)$: $\omega_i(k)=\omega_L$ and
$\omega_i(k)=\omega_L/2$. As it is seen from equations
(\arabic{tenth1}) for all branches of spin waves there exist $k$
which are satisfies such resonance conditions. Vicinities of these
wave vectors are "dangerous" $ $ for the appearance of
instability. From the same formulas it follows that for the
different branches resonance conditions are satisfied with
different values of wave vectors, and therefore each branch can be
considered independently. To find solution nearby the resonance
frequencies we will use the standard procedure, when solution in
the main approximation is sought in a form:
    \begin{eqnarray}
         \label{solut}
         \textbf{X}^{(l)}(t,k^{'},y,z)=a_{i-}(t)\textbf{X}_i(k^{'})\exp(-i\omega_R^{(l)}t+i\textbf{k}^{'}\textbf{r})+\nonumber\\
                       a_{i+}(t)\textbf{X}_i^{*}(k^{'})\exp(i\omega_R^{(l)}t+i\textbf{k}^{'}\textbf{r})+\nonumber\\
                       a_{i-}^{*}(t)\textbf{X}_i^{*}(k^{'})\exp(i\omega_R^{(l)}t-i\textbf{k}^{'}\textbf{r})+\nonumber\\
                       a_{i+}^*(t)\textbf{X}_i(k^{'})\exp(-i\omega_R^{(l)}t-i\textbf{k}^{'}\textbf{r}),
    \end{eqnarray}
$l=1,2$, where $\omega_R^{(l)}$, -~is one of the resonance
frequencies ($\omega_R^{(1)}=\omega_L/2$,
$\omega_R^{(2)}=\omega_L$, we will suppress index $l$ in the
nearest formulas for brevity),$k^{'}$- is wave vector nearby
resonance frequency for the i-mode, $a_{i}(t)$ - are "slowly" $ $
varying functions of time, i.e. $\dot{a}_{\pm}\sim\mu a_{\pm}$.
Terms, that have frequencies differing from $\omega_R$ on integer
multiple of $2\omega_L$ --- $3\omega_R,5\omega_R,7\omega_R$ appear
in the next orders on $\mu$.

Substitution of solution (\ref{solut}) into system
(\ref{V_equation}) gives us:
    \begin{equation}
            \label{q}
        \textbf{X}_a+\textbf{X}_{\omega_R}=\hat{M}(k^{'})\textbf{X}(t,k^{'},y,z)+\hat{V}(\textbf{k}^{'},t)\textbf{X}(t,k^{'},y,z),
    \end{equation}
where
   \begin{eqnarray}
         \textbf{X}_a=\dot{a}_{i-}(t)\textbf{X}_i(k^{'})\exp(-i\omega_Rt+i\textbf{k}^{'}\textbf{r})+\nonumber\\
                       \dot{a}_{i+}(t)\textbf{X}_i^{*}(k^{'})\exp(i\omega_Rt+i\textbf{k}^{'}\textbf{r})+\nonumber\\
                       \dot{a}_{i-}^{*}(t)\textbf{X}_i^{*}(k^{'})\exp(i\omega_Rt-i\textbf{k}^{'}\textbf{r})+\nonumber\\
                       \dot{a}_{i+}^*(t)\textbf{X}_i(k^{'})\exp(-i\omega_Rt-i\textbf{k}^{'}\textbf{r}),
    \end{eqnarray}
    \begin{eqnarray}
         \textbf{X}_{\omega_R}=-i\omega_Ra_{i-}(t)\textbf{X}_i(k^{'})\exp(-i\omega_Rt+i\textbf{k}^{'}\textbf{r})+\nonumber\\
                       i\omega_Ra_{i+}(t)\textbf{X}_i^{*}(k^{'})\exp(i\omega_Rt+i\textbf{k}^{'}\textbf{r})+\nonumber\\
                       i\omega_Ra_{i-}^{*}(t)\textbf{X}_i^{*}(k^{'})\exp(i\omega_Rt-i\textbf{k}^{'}\textbf{r})+\nonumber\\
                       -i\omega_Ra_{i+}^*(t)\textbf{X}_i(k^{'})\exp(-i\omega_Rt-i\textbf{k}^{'}\textbf{r}).
    \end{eqnarray}

Taking into account that $\textbf{X}_i(k^{'})$ and
$\textbf{X}^{*}_i(k^{'})$ are eigenvectors of $\hat{M}$,
corresponding to the frequencies $\omega_i(k^{'})$ and
$-\omega_i(k^{'})$ one can rewrite equation (\ref{q}) as:
\begin{equation}
   \textbf{X}_a+\textbf{X}_{\omega_R}-\textbf{X}_{\omega_k}=\hat{V}(\textbf{k}^{'},t)\textbf{X}(t,k^{'},y,z),
\end{equation}
or
\begin{equation}
   \textbf{X}_a-\frac{\varepsilon(k^{'})}{\omega_R}\textbf{X}_{\omega_R}=\hat{V}(\textbf{k}^{'},t)\textbf{X}(t,k^{'},y,z),
\end{equation}
%\begin{equation}
where   $\varepsilon(k^{'})=\omega(k^{'})-\omega_R$.

Let us multiple the last equation  by $\exp
(-i\textbf{k}^{'}\textbf{r})$ and take integral over volume. As
the result $a_{i-}^{*}$ and $a_{i+}^{*}$ vanish. Expressing
cosines and sines in terms of exponents:
\begin{eqnarray}
        \textbf{X}_a^{(l)}-\frac{\varepsilon^{(l)}(k^{'})}{\omega_R^{(l)}}\textbf{X}^{(l)}_{\omega_R^{(l)}}=(\hat{V}^{(1)}_{+}(\textbf{k}^{'})\exp(2i\omega_R^{(1)}
        t)+\hat{V}^{(1)}_{-}(\textbf{k}^{'})\exp(-2i\omega_R^{(1)}t)+\nonumber\\
        +\hat{V}^{(2)}_{+}(\textbf{k}^{'})\exp(2i\omega_R^{(2)}
        t)+\hat{V}^{(2)}_{-}(\textbf{k}^{'})\exp(-2i\omega_R^{(2)}t))        \mathbf{X}^{(l)}(t),
    \end{eqnarray}
one obtains the sum of terms with different powers of exponents.
Since $V(t)$ contains cosines and sines of $2\omega_R^{(l)} t$,
coefficients $a_{i-}(t)$ and $a_{i+}(t)$ are related by exponents
with the same powers $\pm i\omega_R^{(l)} t$. The resulting
equations are multiplied by $\exp(\pm i\omega_R^{(l)}t)$ and
averaged over rapid oscillations. Finally, after making projection
of equations on eigenvector of i-mode one obtains system of two
differential equations of first order which relates $a_{i-}(t)$
and $a_{i+}(t)$:
\begin{eqnarray}
        \label{slow}
        \dot{a}_{i+}(t)-i\varepsilon^{(l)}(k^{'}){a}_{i+}(t)=\frac{<\textbf{X}^*_i\hat{V}^{(l)}_{-}(\textbf{k}^{'})\textbf{X}^*_i> }{|
        \textbf{X}_i
        |^2}a_{i-}(t),\\
         \dot{a}_{i-}(t)+i\varepsilon^{(l)}(k^{'}){a}_{i-}(t)=\frac{<\textbf{X}_i\hat{V}^{(l)}_{+}(\textbf{k}^{'})\textbf{X}_i> }{|
        \textbf{X}_i
        |^2}a_{i+}(t).
     \end{eqnarray}

System (\ref{slow}) has solution proportional to
$\exp{(\lambda^{(l)} t)}$, where $\lambda^{(l)}$ is defined by:
     \begin{equation}
         \label{increment}
         \lambda^{(l)}_{1,2}=\pm\frac{1}{2}\left(\frac{\displaystyle<\textbf{X}_i
         \displaystyle\hat{V}^{(l)}_{+}(\textbf{k}^{'})
         \textbf{X}_i>\displaystyle<\textbf{X}_i^*
          \displaystyle\hat{V}^{(l)}_{-}(\textbf{k}^{'}) \textbf{X}_i^*>}{\displaystyle|
        \textbf{X}_i\mid^4}-[\varepsilon^{(l)}(k^{'})]^2\right)^\frac{1}{2}.
     \end{equation}

Resonance corresponds to the value of $k^{'}$, when
$\varepsilon^{(l)}(k^{'})=0$. In a region of $k^{'}$ close to
resonance expression in the brackets is positive. Then one of the
values of $\lambda^{(l)}$ corresponds to the growth of amplitude
of oscillations, i.e. development of instability begins.

{\bf 3.} Let us consider all possible cases of resonances. For
each mode we will write: law of dispersion, eigenvector of this
oscillation and increment, which is obtained on the condition of
resonance.

\textbf{First mode.} Law of dispersion:
\begin{equation}
 \omega_1^2=k^2.
\end{equation}

Eigenvector:
$$
X_{1-}(k)=\left(
\begin{array}{c}
0\\
0\\
1\\
\displaystyle
-i\omega_1(k)\cos\beta \\
0\\
\displaystyle i\omega_1(k)(1-\cos\beta)
\end{array}
\right)
$$

Resonance at the frequency $\frac{\omega_L}{2}$:
\begin{equation}
k^{'}=\pm\frac{\omega_L}{2}.
\end{equation}

Increment of growth:
     \begin{equation}
        \label{increment1}
        \lambda_1^{(1)}=\mu\frac{\omega_L}{4}\cdot\frac{\sin\beta|1-2\cos\beta|}{2\cos^2\beta-2\cos\beta+5}\cdot\sin 2\delta,
     \end{equation}
where $\delta$ - is the angle between direction of the wave vector
and z-axis. Maximum increment corresponds to the direction:
      \begin{equation}
       \delta_1^{(0)}=\frac{\pi}{4}.
      \end{equation}
As it is seen from (\ref{increment1}) increment vanishes in the
case of wave vector directed along y-axis.

Resonance at the frequency $\omega_L$:
\begin{equation}
k^{'}=\pm\omega_L.
\end{equation}

In zeroth order approximation on dipole frequency we have:
     \begin{equation}
        \lambda_1^{(2)}=0.
     \end{equation}
Finite increment appears when the  dipole terms are taken into
account in the equations of motion. In this case resonance
condition is satisfied by  the wave vector:
\begin{equation}
k^{'}=\omega_L-\frac{1}{10}(1+4\cos\beta)\frac{\Omega^2}{\omega_L},
\end{equation}
 and eigenvector has corrections of the order of
$\Omega^2/\omega_L^2$. With these corrections the increment will
be equal to:
\begin{equation}
     \label{increment2}
     \lambda_1^{'(2)}=\frac{\mu}{5}\frac{\Omega^2}{\omega_L}
     \sin^2{\beta}\frac{(1+4\cos\beta)^{1/2}}{(1+\cos\beta)^{1/2}}|1-2\cos\beta|\frac{\sin^2\delta}{2+2\cos^2{\beta}-2\cos\beta}
  \end{equation}
  Maximum   increment corresponds to the direction:
      \begin{equation}
       \delta_1^{'(0)}=\frac{\pi}{2}.
      \end{equation}
\textbf{Second mode.}  Law of dispersion:
\begin{equation}
 \omega_2^2=\frac{1}{2}\omega_L^2+k^2-\frac{1}{2}\omega_L\sqrt{\omega_L^2+4k^2}.
\end{equation}

Eigenvector:
$$
X_{2-}=\left(
\begin{array}{c}
1\\
i\sin\beta\cdot(\displaystyle \frac{k^2}{\omega\cdot\omega_2(k)}-\frac{\omega_2(k)}{\omega})\\
1-\cos\beta\\
\displaystyle
-\displaystyle \frac{k^2}{\omega_2(k)}(\sin^2\beta) \\
i\sin\beta\cdot(\displaystyle \frac{k^2}{\omega}-\frac{\omega_2^2(k)}{\omega})\\
-\displaystyle \frac{k^2}{\omega_2(k)}(\sin^2\beta)
\end{array}
\right)
$$

Resonance at the frequency $\frac{\omega_L}{2}$:
\begin{equation}
k^{'}=\pm\frac{\sqrt{3}\omega_L}{2}.
\end{equation}
Increment of growth (to zero order approximation on dipole energy)
     \begin{equation}
        \lambda_2^{(1)}=0.
     \end{equation}
Resonance at the frequency $\omega_L$ (taking into account dipole
energy):
\begin{equation}
k^{'}=\sqrt{2}\omega_L-\frac{2}{15}(1-\cos\beta)\frac{\Omega^2}{\omega_L}.
\end{equation}
Increment:

  \begin{equation}
 \label{increment3}
  \lambda_2^{'(2)}=\frac{2\mu}{5}\frac{\Omega^2}{\omega_L}\sin^2{\beta}(1-\cos\beta)|1-4\sin^2{\beta}|\frac{\sin^2\delta}{12-2\cos\beta-17\cos^2{\beta}+8\cos^4{\beta}},
  \end{equation}
has its maximum for   $\delta_2^{'(0)}=\pi/2.$

\textbf{Third mode.} Law of dispersion:
\begin{equation}
  \omega_3^2=\frac{1}{2}\omega_L^2+k^2+\frac{1}{2}\omega_L\sqrt{\omega_L^2+4k^2}.
\end{equation}

Resonance at the frequency $\frac{\omega_L}{2}$ is not possible
because the frequency of this mode is larger than $\omega_L$ for
all $k$. Resonance at the frequency $\omega_L$  is also not
possible because near  $k=0$ $\lambda$ is imaginary.

\begin{figure}[h]
   \begin{center}

    \resizebox{8cm}{!}{\includegraphics{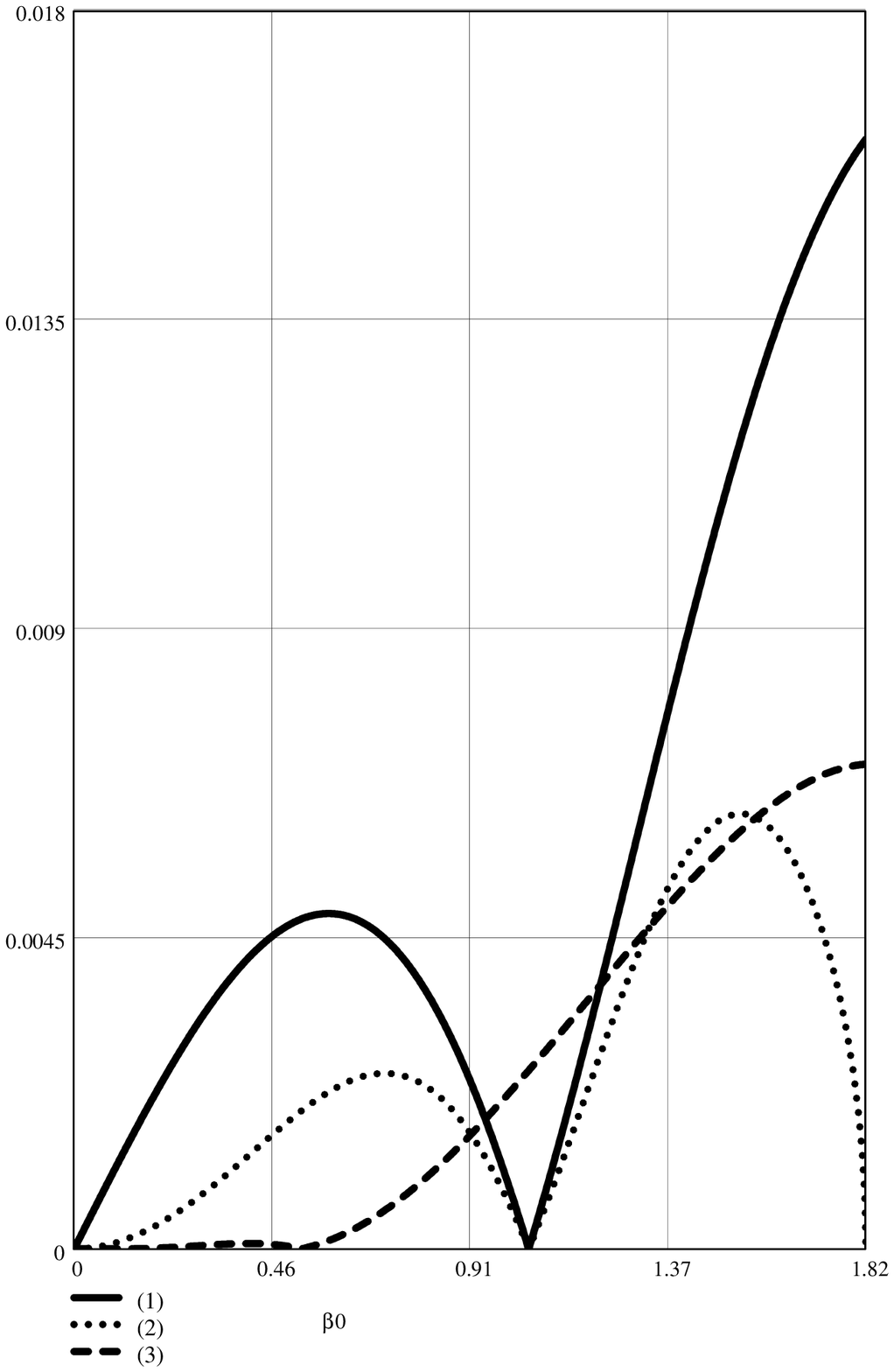}}
      \caption{Dependence of increments for two spin waves modes: (1) - $\lambda_1^{(1)}(\beta^{(0)},\delta_1^{(0)})/\omega_L$ (\ref{increment1}), (2)
- $\lambda_1^{'(2)}(\beta^{(0)},\delta_1^{'(0)})/\omega_L$
(\ref{increment2}),
(3)~-$\lambda_2^{'(2)}(\beta^{(0)},\delta_2^{'(0)})/\omega_L$
(\ref{increment3}), on tipping angle $\beta_0$ in interval
$0\leq\beta<\theta_0=\arccos(-\frac{1}{4})$, $\Omega/\omega_L=1/2$
for (2) and (3).}

      \label{pression}
 \end{center}
 \end{figure}

As it is seen from the Fig. (\ref{pression}) there exists positive
increment at least for  one of the modes of oscillations for all
tipping angles. The main role plays maximum increment that is
found for the first mode in the case of resonance frequency
$\omega_L/2$.

{\bf 4.} The obtained results are correct at $T=0$. At a finite
temperature spin waves damp. This leads to appearance of
temperature threshold of instability. We can take into account
small damping by substituting new complex law of dispersion into
formula for increment (\ref{increment}). Here we should replace
$\varepsilon^{(l)}(k^{'})^2$ by $|\varepsilon^{(l)}(k^{'})|^2$. To
estimate temperature threshold of instability one should use the
law of dispersion with the corrections for damping:
\begin{equation}
     \label{dispersi}
     \omega_i^{'2} (k)=\omega_i^2(k)-2iD(T)\omega_L k^2+O(k^4),
  \end{equation}
where $D(T)$ - is coefficient of diffusion. We substitute
(\ref{dispersi}) into corrected formula (\ref{increment}) with
$k$, which satisfies the resonance condition
$Re(\omega(k))=\omega_R^{(l)}$. After substitution and with
account that $Re(\omega(k))=\omega_R^{(l)}$ we arrive at:
    \begin{equation}
         \lambda^{(l)}(T)=\frac{1}{2}\left(\lambda_{max}^2(T=0)-D^2(T)\frac{\omega_L^2k^4(\omega_R^{(l)})}{(\omega_R^{(l)})^2}\right)^\frac{1}{2}.
     \end{equation}
This formula determines temperature below which instability sets
on:
\begin{equation}
      \label{diffusion_koeff}
      D(T)=\frac{\lambda_{max}^{(1)}\omega_R^{(l)}}{\omega_Lk^2(\omega_R)}.
\end{equation}

We estimate coefficient of diffusion at $T=0.4 T_c$ by using
(\ref{diffusion_koeff}) ($T_c$ - is the temperature of superfluid
transition). For the first mode of oscillations and for the
resonance frequency $\omega_L/2$ at tipping angle $90^{\circ}$ and
for the pressure 20 bar:
\begin{equation}
      D(T)=\frac{2\lambda_{max}^{(1)}c_{\|}^2}{\omega_L^2}=0.027~sm^2/s,
\end{equation}
$c_{\|}=2050\sqrt{1-T/T_c}\approx 1600$ sm/s, $c=3/4$,
$\omega_L=2.9\cdot10^6$ s$^{-1}$. This result can be compared with
the experimental data for the transverse coefficient of diffusion
in $^{3}He-B$ \cite{Experiment}. In the cited work the value of
transverse coefficient of diffusion is approximately equal to 0.03
sm$^2/$s  for the pressure 20 bar and for the Larmor frequency
$2.9\cdot10^6$ s$^{-1}$. Thus the estimated "critical"$~$
coefficient of diffusion is close to the measured one at $0.4T_c$.

{\bf 5.} It follows from the given analysis that at sufficiently
low temperatures homogeneous precession of spin in $^3$He-B is
unstable because of Suhl mechanism. Interaction between precession
and spin waves appears mainly because of the anisotropy of spin
wave velocities. Estimation of decay time of precession as inverse
maximum increment of growth of spin wave amplitude gives result
which does not contradict to the measured value at the lowest
temperatures. Estimation of the temperature threshold of the onset
of  instability using the available data about the value of spin
waves damping falls into the temperature interval in which
transition from stationary precession to catastrophic relaxation
is observed. This allows to consider Suhl instability as the
probable reason of the observed catastrophic relaxation. In order
to make proposed here explanation of catastrophic relaxation
quantitative one should describe more precisely spin waves damping
taking into account direction of propagation. This work is in a
progress.

This work is partly supported by RFBR (grant ¹04-02-16417)
Ministry of Science and Education of the Russian Federation  and
CRDF (grant RUP1-2632-MO04).

\newpage


\begin{thebibliography}{80}
\bibitem{nyeki}  Yu.M. Bunkov, V.V. Dmitriev, Yu.M. Mukharsky et al.,
{\it Europhysics Lett.} {\bf 8}, 645 (1989).
\bibitem{BG} Yu.M. Bunkov, V.L. Golo, J. Low Temp. Phys. {\bf
137}, 625 (2004).
\bibitem{Suhl} H. Suhl, J. Phys. Chem. Solids, {\bf 1}, 209 (1957).
\bibitem{mizus} T. Matsushita, R. Nomura, H.H. Hensley et al., J. Low Temp. Phys. {\bf
105}, 67 (1996).
\bibitem{Ohmi} T. Ohmi, M. Tsubota, J. Low Temp. Phys. {\bf
83}, 177 (1991).
\bibitem{Fomin1} I. A. Fomin, Zh. Exp. Teor. Fiz. {\bf 84}, 2109~(1983) [Sov. Phys. JETP {\bf 57,} 1227~(1983)].
\bibitem{Landau} L. D. Landau and E. M. Lifshitz, {\it Course of Theoretical
Physics,} Vol. 1: {\it Mechanics}, 4th ed. (Nauka, Moscow, 1988,
Pergamon, Oxford, 1989)
\bibitem{Experiment} Yu.M. Bunkov, V.V. Dmitriev, A.V. Markelov et al., Phys. Rev. Lett. {\bf
65}, 867 (1990).
\end{thebibliography}
\end{document}